\documentclass[twocolumn,tighten]{aastex61}

\usepackage{url}
\usepackage{amsmath}

\submitjournal{ApJ}

\shorttitle{Gaia-WISE Extragalactic Astrometric Catalog}
\shortauthors{Paine, Darling, \& Truebenbach}

\begin{document}

\title{The Gaia-WISE extragalactic astrometric catalog}

\correspondingauthor{Jennie Paine}
\email{Jennie.Paine@colorado.edu}

\author{Jennie Paine}
\affiliation{Center for Astrophysics and Space Astronomy \\
Department of Astrophysical and Planetary Sciences \\
University of Colorado, 389 UCB \\
Boulder, CO 80309-0389, USA}

\author{Jeremy Darling}
\affiliation{Center for Astrophysics and Space Astronomy \\
Department of Astrophysical and Planetary Sciences \\
University of Colorado, 389 UCB \\
Boulder, CO 80309-0389, USA}

\author{Alexandra Truebenbach}
\affiliation{Center for Astrophysics and Space Astronomy \\
Department of Astrophysical and Planetary Sciences \\
University of Colorado, 389 UCB \\
Boulder, CO 80309-0389, USA}

\begin{abstract}
The {\it Gaia} mission has detected a large number of active galactic nuclei (AGN) 
and galaxies, but these objects must be identified among the 1000-fold more numerous stars.
Extant astrometric AGN catalogs do not have the uniform 
sky coverage required to detect and characterize the all-sky low-multipole proper motion signals
produced by the barycenter motion, gravitational waves, and cosmological effects.  
To remedy this, we present an all-sky sample of 567,721 AGN in 
{\it Gaia} Data Release 1, selected using WISE two-color criteria. The catalog has fairly uniform sky coverage 
beyond the Galactic plane, with a mean density of 12.8 AGN per square degree. 
The objects have magnitudes ranging from  $G=8.8$ down to {\it Gaia's} magnitude limit, $G=20.7$. 
The catalog is approximately 50\% 
complete but suffers from low stellar contamination, roughly 0.2\%.  We predict that the end-of-mission {\it Gaia} 
proper motions for this catalog will enable detection of the secular aberration drift to high significance 
(23$\sigma$) and will limit the anisotropy of the Hubble expansion to about 2\%.
\end{abstract}


\section{Introduction} \label{sec:intro}

The {\it Gaia} mission will provide astrometric and proper motion measurements for 
a large number of bright active galactic nuclei (AGN), but separating the $\sim10^6$ extragalactic objects
from the $\sim10^9$ stars remains challenging \citep{gaia1}.
Current catalogs include the Large Quasar Astrometric Catalog (LQAC; \citealt{lqac3}), the V\'eron Catalog
of Quasars and AGN \citep{veron}, the \citet{secrest} catalog of mid infrared AGN,
 and the {\it Gaia} Universe Model Snapshot (GUMS), a simulated catalog \citep{robin}.
Many of these catalogs are dominated by the Sloan Digital Sky Survey (SDSS) footprint that covers 
35\% of the sky \citep{sdssdr9}, which is problematic for all-sky proper motion studies that attempt to detect low-multipole
correlated proper motion signals such as the secular aberration drift dipole \citep{titov&lambert, xu, truebenbach2}, the stochastic gravitational 
wave background quadrupole \citep{gwinn1997,titov2011,book2011,darling2017}, 
or the isotropy of the Hubble expansion \citep{darling2014,chang,bengaly}.  

Desirable features of extragalactic proper motion catalogs are all-sky, uniform selection, and
low stellar contamination. Completeness is not very important: it impacts the signal-to-noise
of correlated global proper motions, which scales with the square root of the number
of objects. In this work, we consider only low-multipole proper motion signals, but completeness 
will ultimately determine the maximum multipole that can be 
studied due to the limiting sky density of sources. Stellar contamination is the largest concern for 
detecting global signals of a few $\mu$arcsec yr$^{-1}$ because stellar proper motions can be large
and significant and therefore dominate the individually non-significant extragalactic proper motions.  
What stellar contamination remains in any given extragalactic catalog may be addressed using 
a non-Gaussian permissive likelihood function as described in \citet{darling2017}.  

This paper presents the {\it Gaia}-WISE extragalactic astrometric catalog, a catalog designed to have 
low stellar contamination and fairly uniform sky coverage outside of the Galactic Plane.  
Section \ref{sec:selection} presents the WISE color-color selection used to identify AGN and exclude stars, and
Section \ref{sec:results} explores the sky distribution of the catalog, its optical and mid-IR properties, its
redshift distribution, and the expected end-of-mission proper motion uncertainties.  Section \ref{sec:applications}
predicts the performance of this catalog in detecting the secular aberration drift caused by the barycenter acceleration 
about the Galactic Center.  Section \ref{sec:applications} also predicts the expected {\it Gaia} 
sensitivity to anisotropy in the Hubble expansion.  We discuss the ramifications of this work 
and the future prospects for extragalactic proper motion studies in Sections \ref{sec:discussion} and 
\ref{sec:conclusions}. We assume a Hubble constant of $H_0 = 72$ km s$^{-1}$ Mpc$^{-1}$ and a flat cosmology (other cosmological assumptions are not required).

\section{Catalog Selection Method} \label{sec:selection}

The WISE survey is an all-sky mid-infrared (MIR) survey in the 3.4, 4.6, 12 and 22 $\mu$m bandpasses
(W1, W2, W3, and W4, respectively; \citealt{wise}). The {\it AllWISE} data release, used in this work,
combines data from the cryogenic and post-cryogenic \citep{neowise} survey phases, and provides 
better sensitivity and accuracy over previous WISE data releases. WISE colors have 
been shown to cleanly separate AGN from stars and normal galaxies, and several methods
exist in the literature for selecting AGN with WISE (e.g. \citealt{assef, mateos2012, stern2005, stern2012, truebenbach}). 
To create our catalog of {\it Gaia} AGN, we did not consider selection methods using only a 
W1-W2 color cut in order to avoid contamination from brown dwarfs at low Galactic
latitudes, which can reside in the color space selected by single color cuts \citep{kirkpatrick}. 

We employed the ALLWISE catalog of MIR AGN described in \citet{secrest}. The catalog is based on
the WISE two color selection technique of \citet{mateos2012}
which has cuts in the W1$-$W2 and W2$-$W3 color space, referred to as the 
color wedge. 
This AGN color wedge was defined based on the Bright Ultrahard {\it XMM-Newton}
survey (BUXS), one of the largest flux-limited samples of `ultrahard' X-ray-selected 
AGN, but the method does not employ X-ray selection directly. 
BUXS is comprised of 258 objects, of which 56.2\% are type 1 AGN and nearly the 
rest are type 2. BUXS type 2 AGN are intrinsically less luminous than type 1 AGN.
Since the completeness of the MIR wedge has a strong dependence on luminosity,
the wedge preferentially selects type 1 AGN. 
\citet{secrest} selected 1.4 million MIR AGN using ALLWISE profile fitting magnitudes
with S/N $\ge 5$ and the color wedge criteria of \citet{mateos2012}. They included an additional
 constraint of limiting to ALLWISE sources with 
$\texttt{cc\_flags} = ``0000"$ to avoid sources contaminated by image artifacts.

We cross-matched the \citet{secrest} catalog of MIR AGN with {\it Gaia} Data Release 1
using {\it allwise\_best\_neighbour}, the precomputed WISE cross-match table provided in
the {\it Gaia} archive \citep{marrese}. The table includes only the most likely matches between the 
WISE and {\it Gaia} catalogs, called ``best neighbours." 
Since {\it Gaia} is used as the leading catalog in cross-matching, 
a {\it Gaia} source may be matched to multiple sources from an external catalog. 
\citet{marrese} then determine the best match to the {\it Gaia} source using the
angular distance, position errors, epoch difference, and density of sources in 
the external catalog.
A small number of {\it Gaia} sources have $G>21$, fainter
than {\it Gaia's} nominal magnitude limit of 20.7, which are likely incorrectly determined magnitudes
\citep{gaia1}. Such objects were excluded from the cross-match. Additionally, all stars from
the Tycho 2 survey were removed to avoid stellar contamination, which excluded 65 objects.
We discuss possible further stellar contamination in Section~\ref{sec:contamination}.
The resulting catalog of {\it Gaia} MIR AGN contains 567,721 objects. The first ten objects 
are given in Appendix \ref{app:catalog}, and the full catalog is available online.

\subsection{Completeness} \label{sec:completeness}

The completeness of the WISE color wedge selection is dependent on the ratio of
the AGN luminosity to the host luminosity because host galaxy
light can contaminate the MIR emission \citep{mateos2012,padovani}. Thus, lower luminosity AGN
will have colors of normal galaxies and will be excluded by the color wedge.
To assess the completeness of our catalog, we compared the catalog to the sample of 
SDSS DR9 QSOs \citep{sdssdr9} in {\it Gaia}. 
SDSS QSOs were identified in the {\it Gaia} source catalog via the cross-matching
algorithm provided in the {\it Gaia} archive with a matching radius of 1 arcsecond. 
44.6\% of all {\it Gaia}-SDSS QSOs were also 
identified by the WISE color wedge, suggesting that our sample is missing more than
half of all AGN in the {\it Gaia} catalog. 
Only 49.3\% of {\it Gaia}-SDSS QSOs have S/N $>$ 5 detections and zero contamination and confusion flags in all three WISE bands; 
most of the incompleteness of the {\it Gaia}-WISE catalog is therefore due to
non-detections in the least-sensitive WISE W3 band. 
Among the WISE-detected Gaia-SDSS QSOs, 90.2\% lie in the WISE MIR color wedge.
The remaining quasars generally have bluer W1$-$W2 colors than the color wedge, 
likely due to contamination by host galaxy starlight. 

\subsection{Stellar Contamination}\label{sec:contamination}

\citet{mateos2012} find that contamination by normal galaxies in the MIR wedge is minimal.
For astrometric purposes, however, objects need only be extragalactic, so unresolved galaxies
are acceptable. Contamination by Galactic stars is of much greater 
concern due to their large proper motions. 

To assess any remaining stellar contamination after omitting the Tycho stars, we cross matched our 
sample with the SDSS DR12 catalog \citep{sdssdr12}. 229,073 AGN in our sample reside within
the SDSS footprint, and 65,575
 have a spectroscopic classification from SDSS. Of those, only 104 objects (0.16\%) are identified by their 
spectroscopic classification as stars. Extrapolating to the whole sky gives 
approximately 910 total stars in our sample, suggesting negligible contamination from 
stars. 
We also consider contamination from dusty stars that would not be found in our SDSS
cross-match. \citet{nikutta} find that a majority of objects brighter than W1$=11$ are Galactic 
stars. Our sample contains 1,836 objects with W1$<11$, which indicates a maximum of  0.32\% 
contamination from dusty stars.

\section{Results} \label{sec:results}

\begin{deluxetable*}{cccccccccc} 
\tabletypesize{\scriptsize} 
\tablewidth{0pt} 
\tablecaption{Catalog Statistics \label{tab:stats}} 
\tablehead{
    & \colhead{$G$}  & \colhead{W1} & \colhead{W2} & \colhead{W3} & \colhead{W1$-$W2} & \colhead{W2$-$W3} & \colhead{Redshift} & \colhead{$\sigma_{\mu,RA}$\tablenotemark{a}} & \colhead{$\sigma_{\mu,Dec}$\tablenotemark{a}}\\
    & \colhead{(mag)} & \colhead{(mag)} & \colhead{(mag)} & \colhead{(mag)} & \colhead{(mag)} & \colhead{(mag)} & & \colhead{($\mu$as yr$^{-1}$)} & \colhead{($\mu$as yr$^{-1}$)} }
\startdata 
Mean    	& 	19.3 	& 	15.2 	& 	14.0 	& 	10.9  & 	1.2 	& 	3.0  	& 	1.3 	& 	236  & 	218 \\
Median  	&	19.4 	& 	15.3 	& 	14.1 	& 	11.1   & 	1.2 	& 	3.0 	& 	1.2 	& 	205  & 	191 \\
Minimum 	&  8.8	    & 	4.8  	& 	3.7  	& 	0.2   	& 	0.5 	& 	2.0 	& 	0.0 	& 	2    	& 	3 \\
Maximum 	& 	21.0 	& 	18.8 	& 	17.1 	& 	12.9  & 	2.2 	& 	5.8 	& 	7.0 	& 	1062 & 	797\\
\enddata
\tablenotetext{a}{{\it Gaia} expected end-of-mission proper motion uncertainty (see Section \ref{sec:errors}).}
\end{deluxetable*}

\subsection{Sky Distribution}\label{sec:distribution}

\begin{figure*}[t!]
    \centering
    \plotone{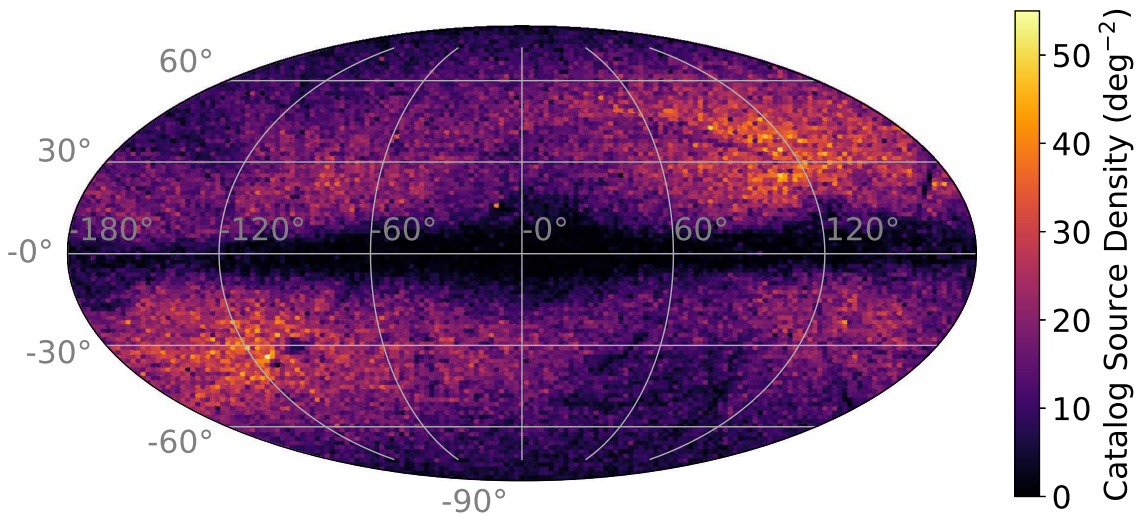}
    \caption{{\it Gaia}-WISE extragalactic astrometric 
      catalog density plot in Galactic coordinates. The colorbar indicates the number of objects per deg$^{2}$.}
    \label{fig:density}
\end{figure*}

Figure~\ref{fig:density} illustrates the distribution of {\it Gaia}-WISE
AGN on the sky. The lower density of AGN at low Galactic latitudes is due to 
a combination of dust along the Galactic plane and the effectiveness of the MIR color wedge 
at excluding stars. Additionally, WISE photometry is limited by confusion near the Galactic 
plane due to high source density \citep{wise}. The higher densities near the ecliptic poles are due to increased
coverage by both WISE and {\it Gaia}. The mean and median densities above the Galactic plane ($b>15^\circ$) 
are 12.8 and 12.0 objects per deg$^{2}$, respectively, and the maximum density is 55 objects per deg$^{2}$. 

\subsection{Optical Properties}\label{sec:optical}

\begin{figure}[ht!]
    \centering
    \includegraphics[width=\columnwidth]{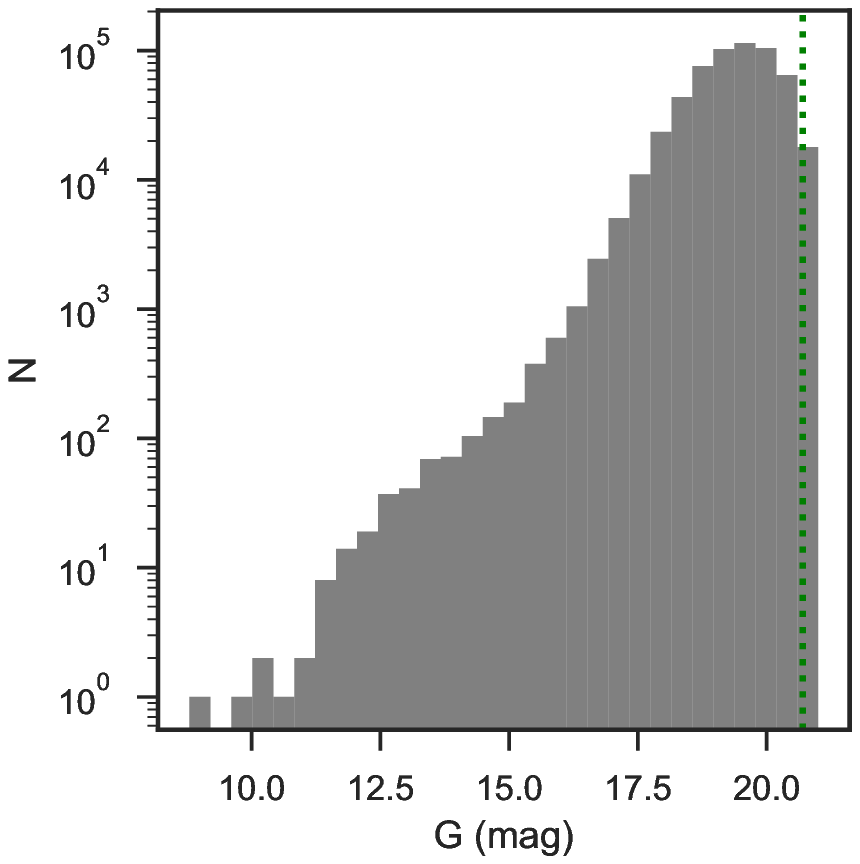}
    \caption{Distribution of {\it Gaia} G-band magnitudes in the {\it Gaia}-WISE extragalactic astrometric catalog. 
    The green dotted line indicates {\it Gaia's} nominal magnitude limit, $G=20.7$.}
    \label{fig:G}
\end{figure}

{\it Gaia} surveys the sky down to $G=20.7$, with a small fraction of objects 
at $G>21$ \citep{gaia1}. As illustrated in Figure~\ref{fig:G}, the majority 
of WISE AGN lie at the fainter end of {\it Gaia's} magnitude distribution. 
Statistics for the distribution of $G$ magnitudes are listed in Table~\ref{tab:stats}.

\subsection{Mid-IR Properties}\label{sec:MIR}
\begin{figure}[ht!]
    \centering
    \includegraphics[width=\columnwidth]{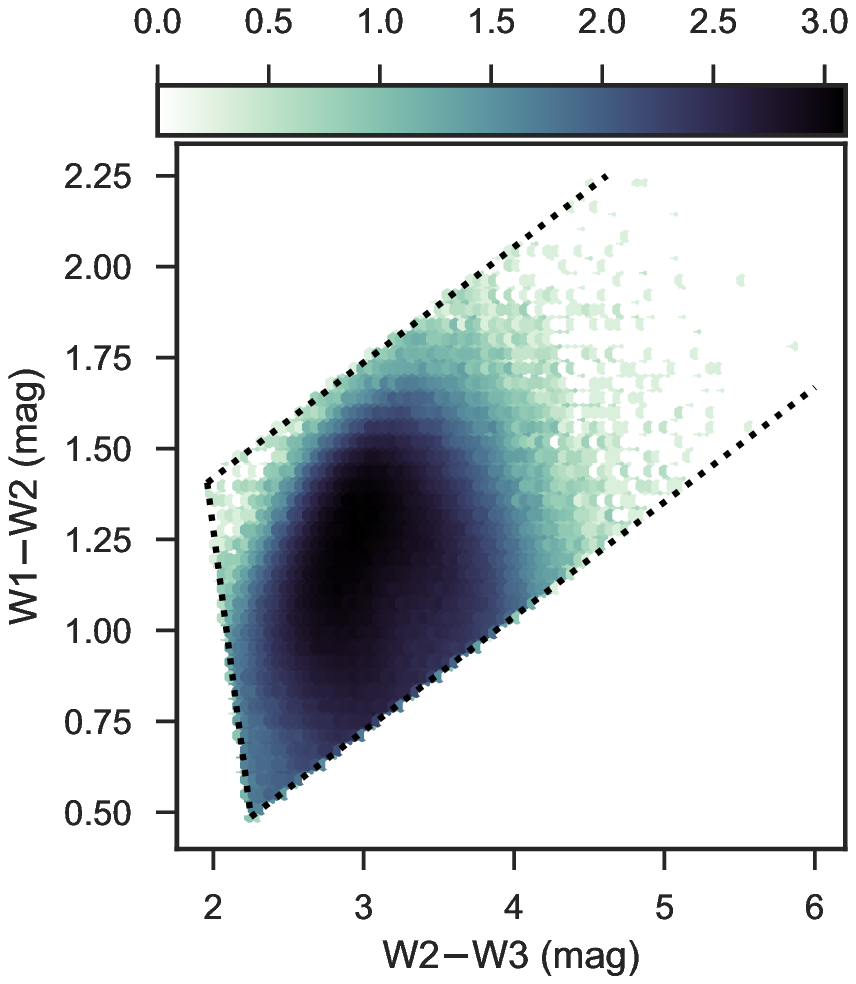}
    \caption{WISE colors for {\it Gaia} MIR AGN. The dashed lines indicate the color 
    wedge of \citet{mateos2012}. The color bar indicates the logarithm of the number of objects per hexagonal bin.}
    \label{fig:colors}
\end{figure}

\begin{figure*}
    \plotone{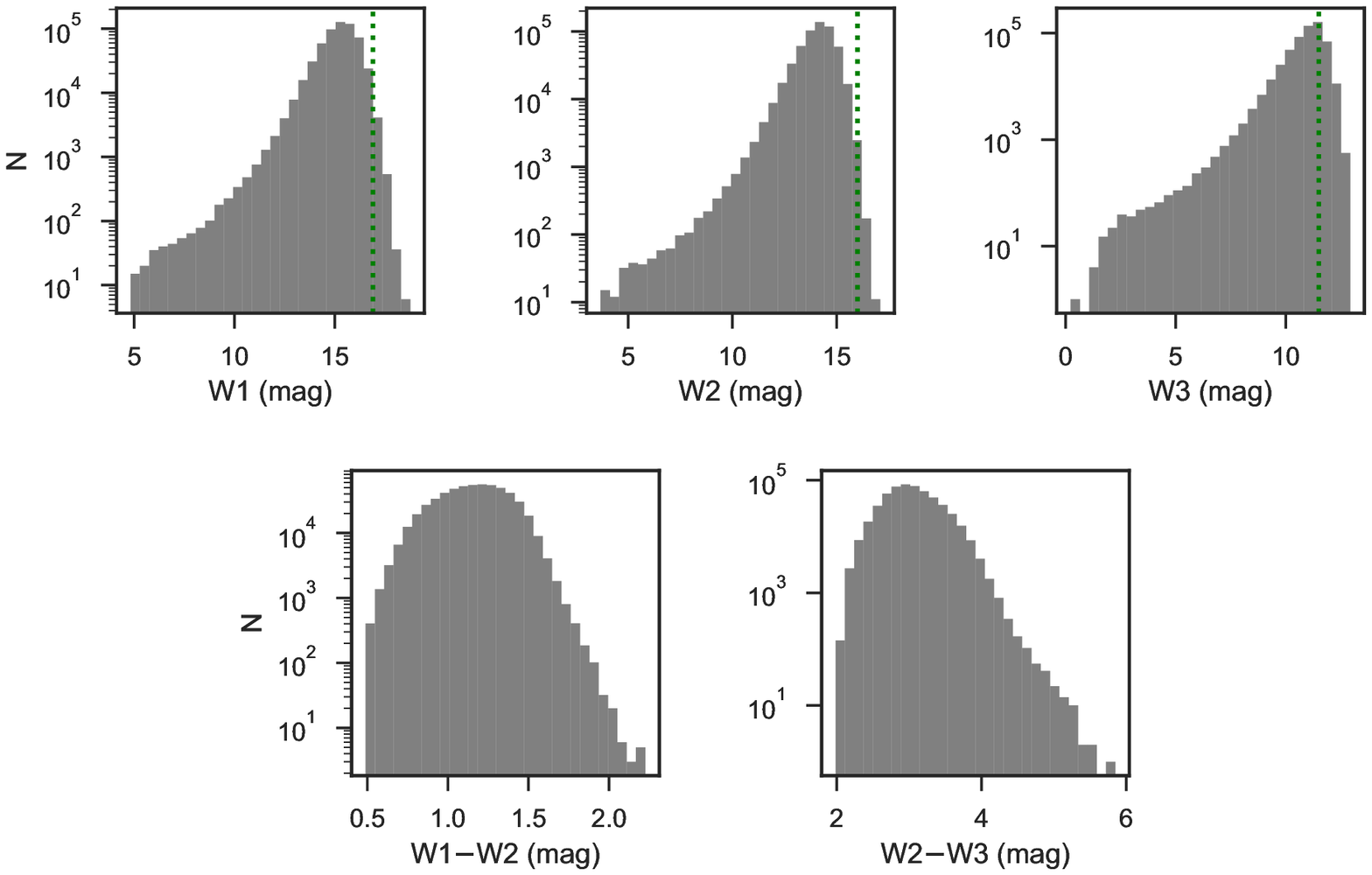}
    \caption{Distribution of W1, W2, and W3 band magnitudes, and W1$-$W2 and W2$-$W3 colors in the {\it Gaia}-WISE extragalactic astrometric catalog. 
    Green dotted lines show the nominal S/N $=$ 5 magnitudes for each band (16.9, 16.0, and 11.5 for W1, W2, and W3, respectively).}
    \label{fig:Wbands}
\end{figure*}

The WISE two color distribution for our catalog is shown in Figure~\ref{fig:colors},
along with the \citet{mateos2012} wedge. 
The majority of objects reside in a locus near the bluer end of the color wedge, 
with a small number of outliers with redder colors. The distribution around the locus tapers before
the color cuts, suggesting that the color wedge captures most of the 
AGN population, except for the bottom right cut where AGN colors 
begin to overlap with the color space occupied by normal galaxies. 
The distributions of WISE W1, W2, and W3 magnitudes, and W1$-$W2 and W2$-$W3 colors are shown in
Figure~\ref{fig:Wbands}; statistics for these distributions are given in Table~\ref{tab:stats}.

\subsection{Redshifts}\label{sec:redshifts}

Redshifts were obtained 
for objects with spectroscopic redshifts from SDSS. Redshifts with nonzero warning flags or
negative errors were discarded, since a negative redshift error indicates a poor fit even if the warning
flag is zero. This yielded redshifts for 90,365 objects ($\sim 15\%$).
Note that this distribution is incomplete and subject to selection bias due to targeted
quasar surveys by SDSS and the redshift sensitivity biases thereof. 
The catalog contains 202 redshifts above $z=4$, which is unexpectedly high considering {\it Gaia's} 
magnitude limit. However, a majority of these are confirmed quasars in the SDSS Baryon Oscillation 
Spectroscopic Survey (BOSS) quasar catalog, of which many were selected for the survey
using WISE colors \citep{paris}.

\begin{figure}[ht!]
    \centering
    \includegraphics[width=\columnwidth]{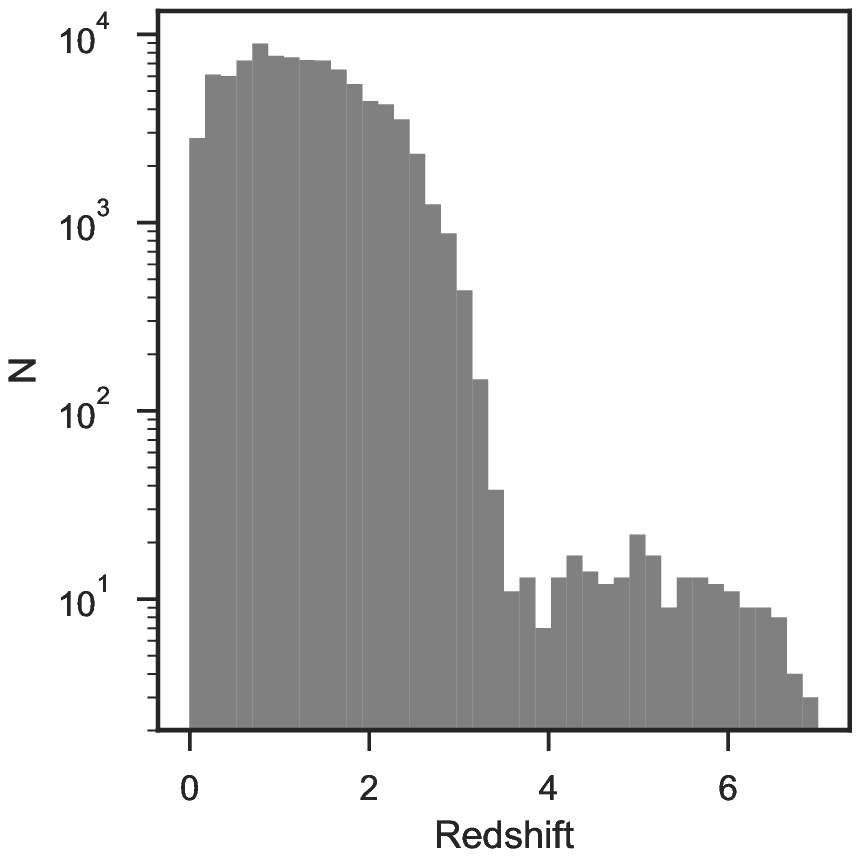}
    \caption{Distribution of redshifts in the {\it Gaia}-WISE extragalactic astrometric catalog, where available (Section \ref{sec:redshifts}). }
    \label{fig:redshift}
\end{figure}

\subsection{Proper Motion Uncertainties}\label{sec:errors}

\begin{figure}[ht!]
    \centering
    \includegraphics[trim={0.75cm 0.5cm 0.75 0.5cm},clip,width=\columnwidth]{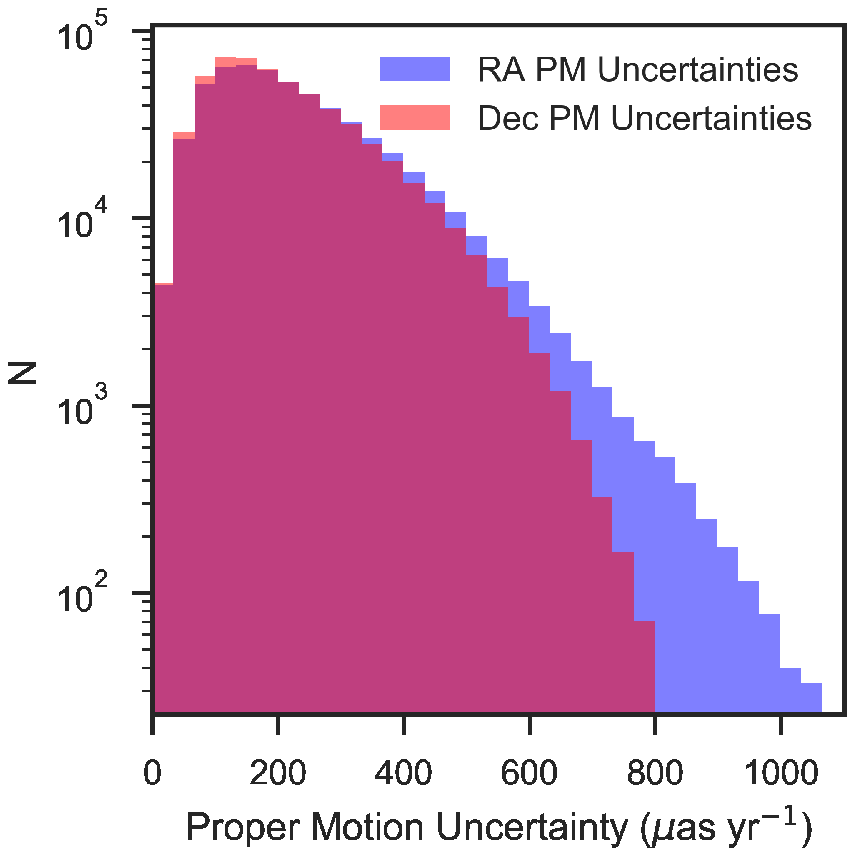}
    \caption{Predicted proper motion uncertainties in both right ascension (blue) and 
    declination (pink), with overlapping values shown in magenta. }
    \label{fig:PM_errors}
\end{figure}

{\it Gaia} DR2 will include positions, 
proper motions, and parallaxes --- or limits on these quantities --- for all objects. Predicted proper motion standard errors can be 
calculated ahead of the release using Gaia performance
characteristics.\footnote{\url{www.cosmos.esa.int/web/gaia/science-performance}} 
The PyGaia Python toolkit is an implementation of Gaia performance models that can be used 
for basic simulation and analysis of Gaia data, including calculation of proper motion uncertainties.
We utilized the PyGaia Python toolkit to calculate predicted proper motion uncertainties for each AGN, 
shown in Figure~\ref{fig:PM_errors}.
This calculation relies on each object's G magnitude, $V-I_C$ color, and ecliptic latitude.
For objects where the $V-I_C$ color was not available,
this value was set to zero, which has a negligible impact on the predicted proper motion
uncertainty. The reported uncertainties include known instrumental effects. 
Statistics for the distributions of predicted uncertainties are given in
Table~\ref{tab:stats}. The uncertainties in right ascension proper motion are 
generally larger than in declination, which is a consequence of the {\it Gaia's} scanning law. 

\section{Applications}\label{sec:applications}
Although proper motions for {\it Gaia} AGN will not be available until DR2,
we can use the predicted uncertainties to test {\it Gaia's} potential
capability to detect or constrain select proper motion signals. 
For this purpose, we generate a null proper motion catalog by randomly selecting 
proper motions consistent with zero based on each object's expected errors 
and assuming Gaussian-distributed errors. 
One can then add proper motion signals to the noisy null catalog to study
the expected sensitivity of the {\it Gaia}-WISE catalog to various
correlated proper motions.  These include the secular aberration drift
(Section~\ref{sec:dipole}), an anisotropic Hubble expansion (Section~\ref{sec:shear}), and a
stochastic long-period gravitational wave background \citep{darling2017}.

\subsection{Secular Aberration Drift}\label{sec:dipole}
The aberration of light is an apparent angular deflection of light rays caused by an observer's velocity across the rays
and the finite speed of light.  Aberration can be caused by the Earth's annual motion or the secular
Solar motion in the Galaxy or with respect to the cosmic microwave background rest frame.  If the 
observer experiences a constant acceleration then the aberration will exhibit a secular drift that manifests as an 
apparent proper motion of objects in a dipole pattern converging toward the acceleration vector direction.

The secular aberration drift caused by the solar system's acceleration toward  the Galactic Center (a consequence of its
orbit) is detectable in extragalactic proper motions as a dipole vector field that resembles an electric field and converges on 
the Galactic Center  (e.g. \citealt{xu, titov&lambert, truebenbach2}).
The expected solar acceleration and corresponding secular aberration drift dipole amplitude can be predicted using
the distance to the Galactic center ($R_0$) and the orbital speed of the Sun ($\Theta_0+V_\odot$), which includes solar motion 
$V_\odot$ in the direction of Galactic rotation $\Theta_0$:  
$ a = (\Theta_0+V_\odot)^2/R_0$ and $ |\vec{\mu}| = a/c$.  
\citet{reid} measured $R_0=8.34 \pm 0.16$ kpc and  $\Theta_0+V_\odot = 255.2\pm5.1$ km s$^{-1}$
from the trigonometric parallaxes and proper motions of masers associated with young massive stars.
These yield an acceleration of $ a = 0.80\pm0.04$~cm~s$^{-1}$~yr$^{-1}$ and 
a dipole amplitude of $|\vec{\mu}| = 5.5\pm0.2$~$\mu$as ~yr$^{-1}$. 

An E-mode vector field dipole painted on the sky, $\mathbf{\vec{V}}_{E1} (\alpha,\delta)$, can be expressed as a $\ell=1$ vector spherical harmonic following the notation of \citet{mignard&klioner}:
\begin{align*}
  \mathbf{\vec{V}}_{E1} (\alpha,\delta) &= 
      \left(s_{11}^{Re}\, {1\over2} \sqrt{3\over\pi}\, \sin\alpha+s_{11}^{Im}\, {1\over2} \sqrt{3\over\pi}\, \cos\alpha\right) \mathbf{\hat{e}}_\alpha  \\
   &  +  \Bigg(s_{10}\, {1\over2}\sqrt{3\over 2\pi}\, \cos\delta + s_{11}^{Re}\, {1\over2} \sqrt{3\over\pi}\, \cos\alpha\sin\delta \\
  &  ~~~~~~-s_{11}^{Im}\, {1\over2} \sqrt{3\over\pi}\, \sin\alpha\sin\delta\Bigg) \mathbf{\hat{e}}_\delta
\end{align*} 
where the coefficients $s_{\ell m}^{Re,Im}$ determine the direction and amplitude of the dipole, $\alpha$ and $\delta$ are the 
right ascension and declination coordinates, and $\mathbf{\hat{e}}_\alpha$ and $\mathbf{\hat{e}}_\delta$ are the unit vectors in those directions. 
In this formalism, the expected E-mode dipole caused by the solar orbit about the Galactic Center ($266.4^\circ$, $-29.0^\circ$) is 
$(s_{10}, s_{11}^{Re}, s_{11}^{Im}) = (-7.71 \pm 0.34, 0.615 \pm 0.027, -9.82 \pm 0.44)$ $\mu$as yr$^{-1}$.
 
In order to predict the {\it Gaia} sensitivity to the secular aberration
drift signal, we assigned a proper motion to each object that is consistent with no proper motion by randomly sampling
its predicted Gaussian proper motion error distribution (Section \ref{sec:errors}).  
Over 1000 random trials, we added the expected secular aberration drift signal to the noisy null proper motions, omitting
the uncertainties in the input dipole, and used a least squares minimization to fit a dipole to the data.   The resulting mean 
of the best fit parameters is 
$(s_{10}, s_{11}^{Re}, s_{11}^{Im}) = (-7.73 \pm 0.48, 0.606 \pm 0.337, -9.79 \pm 0.36)$ $\mu$as yr$^{-1}$, 
consistent with the original input dipole, with mean Z-score of 23. 
We therefore predict that {\it Gaia} will produce the best determination
of the secular aberration drift to date.

\subsection{Anisotropic Cosmic Expansion}\label{sec:shear}
Extragalactic proper motions can test the isotropy of the Hubble 
expansion in the current epoch. If we neglect the peculiar motions of galaxies
caused by density inhomogeneities, an isotropic Hubble expansion produces no extragalactic
proper motions.  In contrast, anisotropic expansion will cause extragalactic objects
to stream toward directions of faster expansion and away from directions with slower expansion. 
All-sky proper motion observations can therefore measure the expansion isotropy and constrain
cosmological models that attempt to explain accelerating expansion without invoking dark
energy, such as Lemaitre-Tolman-Bondi models and Bianchi universes
(e.g. \citealt{amendola}).

\citet{quercellini2009} and \citet{fontanini2009} showed that a triaxial expansion can be described using a Bianchi I 
model, which has the metric
\begin{equation}
  ds^2 = -dt^2+a^2(t)\,dx^{\,2}+b^2(t)\,dy^{\,2}+c^2(t)\,dz^{\,2}.  \label{eqn:metric}
\end{equation}
This metric permits different expansion rates along the three axes: 
  $H_{x} = \dot{a}/a$, $H_{y} = \dot{b}/b$, and $H_{z} = \dot{c}/c$.
The observed Hubble parameter would be $H={d\over dt}(abc)^{1/3}/(abc)^{1/3}$, and the Friedmann-Robertson-Walker
metric is recovered for $a(t)=b(t)=c(t)$.  
The expansion can therefore be characterized by the fractional departure from the isotropic Hubble expansion along the coordinate $i$ using a unitless shear parameter:
\begin{equation}
    \Sigma_{i} = \frac{H_{i,0}}{H_0} - 1.
\end{equation}
The principal shearing axes can be arbitrarily oriented on the sky, and 
\citet{darling2014} showed that the proper motion induced by this anisotropy model can be completely described
by a quadrupolar E-mode vector field.

To test the catalog's potential to constrain anisotropy, we performed 1,000 trials of adding a
randomly generated anisotropy signal to the noisy null proper motions and fitting the anisotropy model to attempt to recreate 
the original input signal. We used the shear equation (Equation A1) of \citet{darling2014} to form these 
artificial anisotropy signals. 
For each trial,  shear terms $\Sigma_{x}$, $\Sigma_{y}$, and $\Sigma_{z}$ were drawn from Gaussian 
distributions with mean of zero and random standard deviation sampled from a uniform distribution between 0 and 0.1. The rotation angles were randomly 
selected from a uniform distribution between 0 and $2\pi$, assuming that there is no preferred 
direction for anisotropy. After the signal is added to the null proper motions, we use a least 
squares minimization to fit the shear equation to the data in an attempt to recover the original signal.

\begin{figure}
    \centering
    \includegraphics[width=\columnwidth]{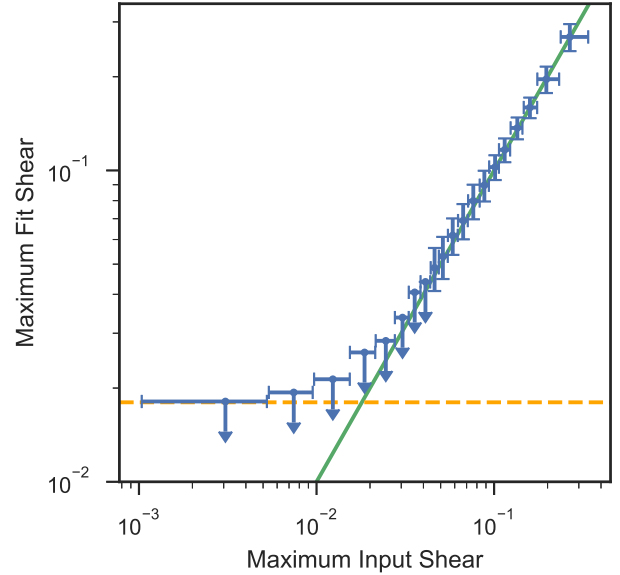}
    \caption{Maximum absolute value of the fit shear vs. the input shear for Hubble expansion anisotropies added to the 
        synthetic {\it Gaia}-WISE AGN catalog proper motions. Non-significant fits are displayed as upper limits. }
    \label{fig:shear_params}
\end{figure}

The shear equation parameters are degenerate due to the rotation degeneracy of the principal axes (no particular axis is required to be the direction of maximum or minimum expansion), and therefore 
individual fit parameters do not necessarily match the original input parameters. 
Instead, we compare the maximum input shear to the maximum fit shear, as shown in 
Figure~\ref{fig:shear_params}. There is a roughly one-to-one correlation for large input values; however, for maximum input shear below $\sim 3 \times 10^{-2}$, noise dominates and the fit parameters tend toward a noise floor of 0.018 (a 1.8\% departure from anisotropy). The fit, however, is not significant for such low input anisotropy. 

\section{Discussion}\label{sec:discussion}

Prior to the first {\it Gaia} data release, the {\it Gaia} Universe model snapshot (GUMS)
simulated a synthetic catalog of objects that {\it Gaia} could have potentially 
observed \citep{robin}. GUMS simulated that nearly one million quasars would be observed 
by {\it Gaia}. Our sample roughly agrees with that number, given that it is about 
50\% incomplete. However, unlike GUMS, our sample consists of real objects actually detected
by {\it Gaia}.

The Large Quasar Astrometric Catalog (LQAC3; \citealt{lqac3}), is a collection of 321,957
objects and represents the complete set of already identified quasars as of 2015. While the LQAC3
reliably contains extragalactic objects, the LQAC3-{\it Gaia} cross-match is dominated 
by the SDSS footprint. Our catalog has a more uniform sky distribution, and is therefore preferable for
the study of low-multipole proper motion signals. 

We expect {\it Gaia}-WISE AGN to be able to measure the secular aberration drift
with 23$\sigma$ significance. 
\citet{mignard} predicted that {\it Gaia} would detect the secular aberration drift with about 
10$\sigma$ accuracy, assuming $10^4$ -- $10^5$ quasars observed by {\it Gaia} with proper 
motion errors lower than predicted here. 
\citet{titov2011} predicted {\it Gaia} to measure the dipole parameters with about 
10\% relative precision. We find that the catalog should be able to measure the dipole parameters 
with higher precision, with the exception of the $s_{11}^{Re}$ component.

While isotropy is a fundamental pillar of cosmology and is well constrained by the cosmic 
microwave background \citep{planck}, {\it Gaia}-WISE AGN will be able to probe the isotropy of 
expansion for the relatively local universe since the majority are at redshift below
2.5 (95th percentile value). 
We predict that {\it Gaia}-WISE AGN will constrain the anisotropy of the Hubble expansion to 
about 2\%. 
\citet{darling2014} showed that the expansion is isotropic to within 7\% in the most 
constrained direction using a catalog of 429 radio sources. 
Local anisotropy has been previously measured using the Hubble parameters derived from Type 1a 
supernovae. \citet{chang} find that the maximum anisotropy of the Hubble parameter is $3\% \pm 1\%$
for a set of supernovae in the redshift range $z<1.4$. \citet{bengaly} find that the maximum variance 
of the Hubble parameter is $(2.30 \pm 0.86)$ km s$^{-1}$ Mpc$^{-1}$ for $z<0.1$, which corresponds
to a maximum departure from isotropy of $3.3\% \pm 1.2\%$.
The {\it Gaia} isotropy measurement will therefore be competitive with and orthogonal to other more traditional methods. 

Our analysis of the astrometric signals that may be detected using {\it Gaia}-WISE AGN has assumed that the proper 
motions of all objects will be determined with the same precision as point sources. In reality, some 
galaxies may appear extended to {\it Gaia}, in which case the precision of the image centroid position 
will be diminished. The intrinsic variability of AGN will be an additional proper motion noise source, 
since variable AGN flux can cause the image centroid to move by up to a few mas for nearby AGN \citep{popovic2012}. 
Microlensing of quasars may also cause the image centroid to shift due to the appearance or disappearance of 
microimages \citep{williams,lewis}. The effect on the centroid position may be as large as tens of $\mu$as due to stellar
mass objects in the lensing galaxy \citep{treyer} or a few mas due to stellar clusters \citep{popovic2013}.
The effects of both AGN variability and microlensing will add uncorrelated noise to the proper motions.
They will therefore be averaged out in the determination of correlated signals such as the secular 
aberration drift and anisotropic expansion, despite adding to the overall noise in the signals.

\section{Conclusions}\label{sec:conclusions}

We presented a catalog of {\it Gaia} AGN selected
using the WISE two color method of \citet{mateos2012}. The catalog contains 
567,721 objects, and we estimate that this sample is
roughly 50\% complete. We find that the WISE wedge reliably selects extragalactic
objects, with only a negligible portion (0.2\%) of our
sample likely contaminated by stars. 
We demonstrated two potential applications of the
catalog, a precise measurement of the secular aberration drift and strong constraints on the isotropy of the Hubble expansion. Based on the expected end-of-mission proper motion uncertainty for each object in the {\it Gaia}-WISE catalog, 
we predict a measurement of the secular aberration drift with 
$\sim 23 \sigma$ significance and a limit on the anisotropy of the Hubble flow of $\sim 2\%$.

\acknowledgments
The authors thank the anonymous referee for helpful feedback.

The authors acknowledge support from the NSF grant AST-1411605 and the NASA grant 14-ATP14-0086.

This work has made use of data from the European Space Agency (ESA)
mission {\it Gaia} (\url{https://www.cosmos.esa.int/gaia}), processed by
the {\it Gaia} Data Processing and Analysis Consortium (DPAC,
\url{https://www.cosmos.esa.int/web/gaia/dpac/consortium}). Funding
for the DPAC has been provided by national institutions, in particular
the institutions participating in the {\it Gaia} Multilateral Agreement.

This publication makes use of data products from the Wide-field Infrared Survey Explorer,
which is a joint project of the University of California, Los Angeles, and the Jet Propulsion
Laboratory/California Institute of Technology, funded by the National Aeronautics and Space
Administration.

Funding for SDSS-III has been provided by the Alfred P. Sloan Foundation, the Participating Institutions, the National Science Foundation, and the U.S. Department of Energy Office of Science. The SDSS-III web site is \url{http://www.sdss3.org/}.
SDSS-III is managed by the Astrophysical Research Consortium for the Participating Institutions of the SDSS-III Collaboration including the University of Arizona, the Brazilian Participation Group, Brookhaven National Laboratory, Carnegie Mellon University, University of Florida, the French Participation Group, the German Participation Group, Harvard University, the Instituto de Astrofisica de Canarias, the Michigan State/Notre Dame/JINA Participation Group, Johns Hopkins University, Lawrence Berkeley National Laboratory, Max Planck Institute for Astrophysics, Max Planck Institute for Extraterrestrial Physics, New Mexico State University, New York University, Ohio State University, Pennsylvania State University, University of Portsmouth, Princeton University, the Spanish Participation Group, University of Tokyo, University of Utah, Vanderbilt University, University of Virginia, University of Washington, and Yale University.

This research has made use of the NASA/IPAC Extragalactic Database (NED) which is operated by the Jet Propulsion Laboratory, California Institute of Technology, under contract with the National Aeronautics and Space Administration.

\software{astropy \citep{astropy}, pyGaia, STILTS \citep{stilts}, TOPCAT \citep{topcat}
          }

\appendix

\section{Catalog}\label{app:catalog}

Table~\ref{tab:catalog} lists the first ten rows of the {\it Gaia}-WISE extragalactic catalog. 
The full catalog containing 567,721 objects will be available as a machine-readable table provided by the publisher.

\begin{turnpage}
\begin{deluxetable}{rcrcrcp{1cm}lrlrlrccc}
\tabletypesize{\scriptsize} 
\tablewidth{0pt} 
\tablecaption{{\it Gaia}-WISE Extragalactic Catalog \label{tab:catalog}} 
\tablehead{
    \colhead{{\it Gaia} ID} & \colhead{RA} & \colhead{$\sigma_{RA}$} & \colhead{Dec} & \colhead{$\sigma_{Dec}$} 
    & \colhead{$G$} & \colhead{{\it ALLWISE} ID} & \colhead{W1} & \colhead{$\sigma_{W1}$} & \colhead{W2} & \colhead{$\sigma_{W2}$} & \colhead{W3} & \colhead{$\sigma_{W3}$} & \colhead{Redshift} & \multicolumn{2}{c}{Proper Motion Uncertainties\tablenotemark{a}} \\ \cline{15-16}
     & \colhead{J2000} & \colhead{} & \colhead{J2000} & & & & & & & & & 
     & &  \colhead{$\sigma_{\mu,RA}$} & \colhead{$\sigma_{\mu,Dec}$} \\
     & \colhead{(degrees)} & \colhead{(mas)} & \colhead{(degrees)} & \colhead{(mas)} & \colhead{(mag)} && \colhead{(mag)} &\colhead{(mag)} & \colhead{(mag)} &\colhead{(mag)} & \colhead{(mag)} &\colhead{(mag)} && \colhead{($\mu$as yr$^{-1}$)} & \colhead{($\mu$as yr$^{-1}$)}
    }
\startdata 
4990063153917291776 & 0.00026196 & 0.4 & -47.64309208 & 0.4 & 18.637 & J000000.06-473835.1 & 14.086 & 0.027 & 13.233 & 0.028 & 9.987 & 0.048 &  & 81 & 81\\
2875546163053982464 & 0.00062956 & 2.6 & 35.51784342 & 1.0 & 18.537 & J000000.15+ 353104.1 & 14.522 & 0.030 & 13.372 & 0.031 & 10.663 & 0.102 &  & 108 & 108\\
2341836724939897216 & 0.00066058 & 0.3 & -20.07434420 & 0.3 & 17.910 & J000000.15-200427.7 & 13.548 & 0.026 & 12.539 & 0.025 & 9.727 & 0.053 &  & 85 & 85\\
4635686437412067840 & 0.00102928 & 1.2 & -78.53449449 & 1.4 & 20.226 & J000000.23-783204.1 & 15.212 & 0.031 & 13.694 & 0.028 & 10.388 & 0.055 &  & 336 & 336\\
2305851255551067776 & 0.00142474 & 3.9 & -41.49299774 & 0.6 & 18.597 & J000000.33-412934.9 & 15.083 & 0.033 & 13.881 & 0.035 & 10.396 & 0.060 &  & 93 & 93\\
2747188660230483712 & 0.00191760 & 0.4 & 9.38565564 & 0.2 & 18.234 & J000000.46+ 092308.2 & 15.316 & 0.042 & 14.019 & 0.044 & 10.518 & 0.108 &  & 113 & 113\\
2420718231737082368 & 0.00308067 & 1.2 & -13.95693841 & 1.0 & 19.833 & J000000.73-135724.8 & 15.894 & 0.053 & 14.556 & 0.058 & 11.170 & 0.147 &  & 371 & 371\\
2341416058663072000 & 0.00345683 & 0.4 & -21.29793756 & 0.4 & 18.551 & J000000.82-211752.5 & 14.668 & 0.031 & 13.405 & 0.032 & 10.934 & 0.130 &  & 132 & 132\\
2744944385199380480 & 0.00408179 & 1.3 & 4.82979136 & 0.4 & 19.661 & J000000.98+ 044947.1 & 15.503 & 0.044 & 13.987 & 0.044 & 10.764 & 0.112 & 1.62 & 338 & 338\\
2746747137592463872 & 0.00424303 & 1.8 & 8.07294561 & 0.7 & 20.003 & J000001.02+ 080422.6 & 15.332 & 0.042 & 14.160 & 0.045 & 11.118 & 0.171 &  & 441 & 441\\
\enddata
\tablenotetext{a}{{\it Gaia} expected end-of-mission proper motion uncertainty (see Section \ref{sec:errors}).}
\end{deluxetable}

\end{turnpage}

\end{document}